%% file: main.tex
\documentclass[alpha-refs]{mywiley}
\usepackage{ifthen,graphicx,lineno}
\usepackage{doi}
\usepackage{subcaption}
\newboolean{@draft}
\newboolean{@annotate}
\setboolean{@draft}{false}
\setboolean{@annotate}{false}
\input{macrofile}
\papertype{Original Article}
\author[1,2]{Jochen Bröcker}
\author[1,2,3]{Eviatar Bach}
\affil[1]{Department of Meteorology and Department of Mathematics and Statistics, University of Reading, Reading, UK}
\affil[2]{Centre for the Mathematics of Planet Earth, University of Reading, Reading, UK}
\affil[3]{National Centre for Earth Observation, Reading, UK}
\corraddress{Jochen Br\"{o}cker and Eviatar Bach, Department of Mathematics and Statistics, University\ of Reading, Whiteknights, Reading, Berkshire, RG6 6AX}
\corremail{j.broecker@reading.ac.uk, eviatarbach@protonmail.com}
\title{A generalisation of the signal-to-noise ratio using proper scoring rules}
\begin{document}


%
\maketitle
\begin{abstract}
A generalised concept of the signal-to-noise ratio (or equivalently the ratio of predictable components, or $\RPC$) is provided, based on proper scoring rules.
This definition is the natural generalisation of the classical $\RPC$, yet it allows one to define and analyse the signal-to-noise properties of any type of forecast that is amenable to scoring, thus drastically widening the applicability of these concepts.

The methodology is illustrated through numerical examples of ensemble forecasts, scored using the continuous ranked probability score (CRPS), and of probability forecasts of a binary event, scored using the logarithmic score. Numerical examples are carried out using both synthetic data with prescribed signal-to-noise ratios as well as seasonal ensemble hindcasts of the North Atlantic Oscillation (NAO) index. The latter have previously been interpreted as having a signal-to-noise ``paradox'', or anomalous signal-to-noise ratio, using the RPC statistic.

For the synthetic data, the $\RPC$ statistic as well as the scoring rule--based ones agree regarding which data sets exhibit anomalous signal-to-noise ratios, but exhibit different variance, indicating different statistical properties. For the NAO data, on the other hand, the different statistics are more equivocal on whether the signal-to-noise ratio is anomalous.
\end{abstract}
\section{Introduction}
The signal-to-noise ``paradox'' in climate forecasting was first described in \citet{kumar_finite_2009,eade_seasonal--decadal_2014,scaife_skillful_2014} in the context of ensemble forecasts on seasonal-to-decadal timescales, and the first use of the term was in \citet{dunstone_skilful_2016}. It is reviewed in \citet{scaife_signal--noise_2018,weisheimer_signal--noise_2024}. It has also been studied on subseasonal timescales \citep{roberts_ensemble_2025}.
It refers to a situation in which the correlation between the ensemble mean and actual verification is larger than the correlation between the ensemble mean and individual ensemble members.
That the ensemble mean is better able to predict the verification than the ensemble members might indeed seem paradoxical.

The signal-to-noise ratio ($\SNR$) as a statistical diagnostic is not new.
To our knowledge it was first introduced to the climate literature in \citet{madden_estimates_1976}.
The idea of estimating $\SNR$s from ensembles by partitioning the ensemble spread into ``signal'' and ``noise'' was considered by various authors and is reviewed in \citet{rowell_assessing_1998}.
Perspectives on the phenomenon have ranged from the dynamical \cite{falkena_detection_2022,dunstone_skilful_2023} to the statistical \cite{brocker_statistical_2023,brener_hypothesis_2024,mahmood_perfect-model_2025}; see \citet{weisheimer_signal--noise_2024} for an overview.
Following \citet{scaife_skillful_2014,brocker_statistical_2023}, we will henceforth refer to such a situation as an {\em anomalous} signal-to-noise ratio.
Although estimated $\SNR$s can indeed be used to diagnose normal or anomalous signal-to-noise ratios, there are alternative (yet equivalent) diagnostics.
The {\em ratio of predictable components}~($\RPC$) statistic, for instance, was introduced in \citet{eade_seasonal--decadal_2014}.

Despite the widespread use of the $\RPC$, a thorough analysis of the statistical properties of the $\RPC$ or equivalent diagnostics is still outstanding.
It is not clear, for instance, when the $\RPC$ of a given data set (or more precisely an estimator of the $\RPC$) is ``significantly'' larger than one; a proper testing methodology would have to be able to answer that question.
The present paper aims to contribute to the development of this methodology.
The precise motivation for this paper is threefold.
Firstly, we want to widen the spectrum of possible statistics for diagnosing an anomalous signal-to-noise ratio.
A greater flexibility in choosing test statistics is likely to be beneficial, rather than just relying on the classical $\RPC$. In particular, other test statistics may have better sampling properties or more accurately capture the probabilistic information contained in the forecasts.
Secondly, the classical $\RPC$ is only applicable to real-valued verifications.
This is a serious limitation, since there are various other types of verifications such as spatial or categorical ones, which as of yet cannot be analysed in terms of the $\RPC$.
Thirdly, by placing the $\RPC$ in a more general information-theoretic framework, we hope to illuminate its theoretical origin and assumptions involved in its use.
We propose to generalise the $\RPC$ by considering instead the ratio of two {\em skill scores}, namely the skill score of the forecast against a verification drawn from the forecast, versus the skill score of the perfectly recalibrated forecast against the actual verifications.
In common with previous authors, {\em skill score} in this paper refers to the score of a forecast, relative to the score of a reference forecast, typically a climatological forecast.
To score forecasts we use proper scoring rules~\citep{broecker06-3}, as these provide a coherent framework for forecast evaluation. See also~\citet{dawid_geometry_2007} for the scoring rule--based notion of entropy, and \citet{takaya_information-based_2025} for its use in a climate context.
In Section~\ref{sec:ratio_skill_scores} we introduce the ratio of skill scores ($\RSS$), the proposed generalisation of the $\RPC$, after we briefly recalled the notion of proper scoring rules and their properties in Subsection~\ref{sec:scoring_rules}.
The $\RSS$ can be applied to any type of forecast that is amenable to scoring.
In Subsection~\ref{sec:classical_snr} we demonstrate that this definition is the natural generalisation of the classical $\RPC$.
In Section~\ref{sec:RPC_interpretation} we briefly discuss some implications of an anomalous signal-to-noise ratio (these apply likewise to the general and the classical definition).
Numerical examples are discussed in Section~\ref{sec:numerical_experiments}.
We consider ensemble forecasts, scored using the continuous ranked probability score~(CRPS), and probability forecasts of a binary event, scored using the logarithmic score. 
We compare the $\RSS$s with the classical $\RPC$ in the context of synthetic data sets as well as data from an operational seasonal forecasting system.
Section~\ref{sec:conclusions} gives concluding remarks and areas for future work.
\section{Mathematical framework}
\label{sec:ratio_skill_scores}
\subsection{Proper scoring rules}
\label{sec:scoring_rules}
Let $E$ be some space that will host the {\em verification, that is, the variable to be forecast.
For the purpose of this paper, $E$ can be taken as either the Euclidean space $\R^d$ or simply a finite set $\{e_1, \ldots, e_K\}$.}
Further, let $\mu, \nu$ be probability distributions over $E$ which the reader might think of as densities (in the case of $E = \R^d$) or probability vectors (of dimension $K$ in the case of $E = \{e_1, \ldots, e_K\}$).
In the geosciences, probability distributions are often represented as ensembles.
Although ensembles provide an accurate representation of the underlying distribution function only in the limit of infinitely many ensemble members, the reader may think of our distributions as ensembles with arbitrarily many members.
A {\em scoring rule} over $E$ is a function $s(\mu, y)$ taking as input such a probability distribution $\mu$, and some $y \in E$ which is a possible value for the verification.
The value $s(\mu, y)$ measures how well $\mu$ forecasts $y$. We adopt the convention of {\em negatively oriented} scores: a smaller score indicates a better forecast.

For any scoring rule $s$ and probability distributions $\mu, \nu$ we define the associated \emph{expected score}
\begin{equation}
        \cS(\mu, \nu) := \int s(\mu, y) \nu(y) \idd y, \label{eq:1.10.1}
\end{equation}
which is the limiting quantity we would obtain by repeatedly drawing a verification from a distribution $\nu$, scoring it against the distribution $\mu$, and then averaging over the draws from $\nu$.
In case $\nu$ is given by an ensemble, the integral in Equation~\eqref{eq:1.10.1} would simply be an average over the ensemble members.
We may draw the verification from $\mu$ as well (i.e.,\ set $\nu$ equal to $\mu$ in this experiment).
This particular form of the expected score we refer to as the {\em entropy} of $\mu$
\begin{equation}
         \cE(\mu) := \cS(\mu, \mu)\label{eq:1.10.2}.
\end{equation}
The entropy measures the average ability of a probability distribution to forecast verifications that are in fact drawn from that same distribution.
The difficulty of this task depends on the shape of the distribution and is evidently a lot easier for a distribution that is, for instance, strongly concentrated around a specific point in $E$ than for a distribution that is broadly spread out. 
The entropy of a distribution can therefore be considered as a measure of how ``random'' that distribution is, thus being a generalisation of the spread of the distribution.
In the context of this formalism, it seems reasonable to require that when it comes to forecasting draws from a certain distribution $\nu$, the distribution $\nu$ itself should be best at this task, on average.
Any other distribution $\mu$ should be worse, or at best equal, in terms of the average score.
A scoring rule $s$ with this property is called {\em proper}.
To define propriety more formally (and for later use) we define the {\em divergence} associated with $s$ through the formula
\beq{eq:1.10.3}
\cD(\mu, \nu) := \cS(\mu, \nu) - \cE(\nu).
\eeq
The scoring rule is {\em proper} if, for all $\mu$, $\nu$, $\cD(\mu, \nu) \geq 0$, and {\em strictly proper} if $\cD(\mu, \nu) = 0$ implies $\mu = \nu$.
For more on scoring rules and the importance of propriety when interpreting scoring rules as measures of forecast performance, see~\citet{broecker06-3,gneiting_strictly_2007}.
We note that the scoring rule being strictly proper implies that $\cD$ is in fact a statistical divergence \citep{gneiting_strictly_2007,bach_inverse_2024}, so it can be regarded as a sort of distance or discrepancy between two probability distributions, although $\cD$ is not necessarily symmetric in its arguments.
Examples of proper scoring rules will be encountered later in Section~\ref{ssec:experimental_setup}.
We henceforth fix a proper scoring rule $s$.

In a typical forecasting problem, we do not see repeated verifications drawn from a fixed distribution and scored against a fixed forecast.
Rather, the forecast itself changes as the forecaster receives and processes information about the future verification.
A {\em probability forecast} is a probability distribution $f$ on $E$ (again the reader might think of $f$ as an ensemble) which at the same time depends on some random variable $\omega$; this dependence reflects the fact that forecasts typically incorporate observations, through data assimilation, that have been collected in the past, and those observations are contaminated by random error.
Those data can be thought of as a random variable $\omega$ in a potentially very high-dimensional space. 
We could indicate this dependence on $\omega$ with a subscript $f_{\omega}$, although we will use this notation sparingly to avoid clutter.
We will also assume the verification $Y_{\omega}$ to be a random variable with values in $E$.
It might seem impractical that we let $Y_{\omega}$ and $f_{\omega}$ depend on the same random variable $\omega$, but it turns out to be more convenient to collate all ``randomness'' in a single variable. 
This does not mean that $Y_{\omega}$ and $f_{\omega}$ contain the same information, since they may depend on different components of $\omega$.
Furthermore, $\omega$ will have some distribution, but we have no need to write it explicitly.
Integrals over the distribution of $\omega$ will be denoted by $\E(\ldots)$; that is, as expectation values.
Given a verification $Y_{\omega}$ and a probability forecast $f_{\omega}$ we define another probability forecast $\pi_{\omega}$, the {\em recalibration} of $f$, as the conditional distribution of $Y$ given $f$, namely
\beq{eq:1.20}
        \pi := \cL(Y | f),
\eeq
where here $\cL$ stands for ``law'', in this case the conditional law of $Y$ given $f$.

If $E$ is just a finite set of alternatives, then we are conditioning on a finite-dimensional vector with random entries.
If in fact $E$ contains just two possible outcomes (say $0$~and $1$), then we are conditioning on a single random number, namely the forecast probability of the event $Y=1$. Readers familiar with the reliability diagram~\cite{broecker06-4} will realise that in this case, the conditional probability distribution $\pi$ corresponds to the perfect recalibration curve. Conditioning on a more general random object (if $f$ is a continuous probability distribution) is possible using the common mathematical definition of the conditional expectation~\citep[see, e.g.,][Ch.4]{breiman_probability_1992}, although it cannot always be done using the familiar Bayes rule.

Note that $\pi$ is a probability forecast over $E$ as well; it is the forecast of $Y$ that optimally uses the information contained in $f$.
For a discussion of the recalibration $\pi$ of $f$ in the context of ensemble forecasts, see~\citet{broecker_assessing_ensembles_serial_dependence_2018,broecker_stratified_serial_dependence_2020}.
We furthermore define $\meanpi := \E(\pi)$, and likewise $\meanf := \E(f)$; both $\meanf$ and $\meanpi$ are probability distributions over $E$, but they do not depend on $\omega$ and therefore are not random. $\meanpi$ can be considered the \emph{climatology} of $Y$---that is, the unconditional distribution from which the verification comes---since
\beq{}
\meanpi = \E(\cL(Y | f)) = \cL(Y).
\eeq
Since $Y$ is a random variable (we could write $Y_{\omega}$), then so is $s(f, Y)$, and if $f$ or $\pi$ are random probability forecasts, then  $\cS(f, \pi), \cE(\pi), \cD(f, \pi)$, etc.,\ are also random, so it makes sense to consider expectations of these quantities.
Between these quantities there exists an important relationship that decomposes the expected score $\E(s(f, Y))$ (where, as opposed to Equation~\eqref{eq:1.10.1}, the expectation is now taken with respect to $\omega$) into three interpretable terms:
\beq{eq:1.30}
        \E(s(f, Y)) = \cE(\meanpi) 
        - \E(\cD(\meanpi, \pi))
        + \E( \cD(f, \pi) ).
\eeq
The three terms on the right-hand side are called, respectively, the {\em entropy} of $\meanpi$, the {\em resolution}, and the {\em reliability} of $f$ (although it must be kept in mind that the definition of these terms refers not only to $f$ but also to $Y$).
It follows from the propriety of the scoring rule $s$ that all terms involving $\cD$ are non-negative (although note that the resolution term carries a minus~sign).
The reliability term is zero if $f = \pi$, and in general it quantifies the extent to which this fails to be true on average, or in other words, by how much $f$ agrees with its optimal recalibration $\pi$.
The stronger the agreement, the smaller the reliability term, and the better the overall score.
The resolution term is zero if $\meanpi = \pi$, and in general it quantifies how much $\pi$ diverges from its average $\meanpi$.
More divergence means a larger resolution term and a {\em better} overall score.
For a proof of Equation~\eqref{eq:1.30} and more details regarding the interpretation, see~\citet{broecker09}.
We note also that
\beq{eq:1.25}
        \E(s(f, Y)) = \E(\cS(f, \pi));
\eeq
see Appendix~\ref{sec:appendix1} for the proof.
The final object we need to introduce is the {\em self-score} of a forecast $f$.
Given a probability forecast $f$, it makes sense to score it against a hypothetical or synthetic verification $Z$ which is drawn from the forecast $f$ itself; that is, $Z\sim f$. 
The resulting score will be referred to as the {\em self-score} of $f$.
Assuming such synthetic verifications, $f$ will be equal to its own recalibration as the relation~\eqref{eq:1.20} holds in the form $f = \cL(Z| f)$; we have defined $Z$ so as to render this true.
The score $\E(s(f, Z))$ can be decomposed as in Equation~\eqref{eq:1.30}, but the reliability term will be zero, so Equation~\eqref{eq:1.30} holds in the form
\beq{eq:1.40}
\E(s(f, Z)) = \E(\cE(f)) = \cE(\meanf) - \E(\cD(\meanf, f)),
\eeq
where the first equality follows from Equation~\eqref{eq:1.25}. Lastly we note, also from Equation~\eqref{eq:1.25}, that
\beq{eq:1.39}
\E(s(\pi, Y)) = \E(\cE(\pi)).
\eeq
With these preparations, we make the following definition.
\begin{defnt}\label{def:snp}
Given a probability forecast $f$, the quantity
\beq{eq:1.40.3}
    \SSC(f) := \frac{\E \cE(f)}{\cE(\meanf)},
\eeq
is called the self--skill~score of $f$.
Given, in addition, a verification $Y$, we say the forecast $f$ exhibits an {\em anomalous signal-to-noise ratio} (with respect to the verification $Y$ and the scoring rule $s$) if $\SSC(\pi) < \SSC(f)$.

Equivalently, we define the \emph{ratio of skill scores} for scoring rule $s$ as
\begin{equation}\label{eq:rss_s}
    \RSS_s = \frac{\SSC(f)}{\SSC(\pi)},
\end{equation}
and say that the forecast exhibits an anomalous signal-to-noise ratio with respect to verification $Y$ and scoring rule $s$ if $\RSS_s > 1$.
\end{defnt}
We note that, using Equations~\eqref{eq:1.40} and \eqref{eq:1.39}, the $\RSS$ can also be written as
\beq{eq:1.41}
    \RSS_s = \left.\bigg(\frac{\E(s(f, Z))}{\cE(\meanf)}\bigg)\middle/\bigg(\frac{\E(s(\pi, Y))}{\cE(\meanpi)}\bigg)\right.,
\eeq
where $Z\sim f$ as before.
The self--skill~score $\SSC(\pi)$ is the ratio of the self-score of $\pi$ and the self-score of the climatology $\meanpi$.
Therefore, the self--skill~score $\SSC(\pi)$ is a {\em relative} measure of performance.
(The term ``skill score'' typically refers to a score relative to some benchmark, but different from many authors, our skill scores are negatively oriented so that a lower self--skill score indicates a higher skill.)
The interpretation of $\SSC(f)$ is that it refers to a situation in which the actual verifications are replaced with synthetic verifications drawn from the forecast $f$ itself.
%
%
%
Therefore, the $\RSS$ is the ratio between the ``anticipated'' skill score of $f$ that would materialise if verifications $Y$ were drawn from $f$ and the ``actually achievable'' skill score of $f$ (through proper recalibration).
An anomalous signal-to-noise ratio thus means that a better skill score can be achieved (at least in theory) through proper recalibration than through taking the forecast at face value.
We will continue this discussion in Section~\ref{sec:RPC_interpretation}.
\subsection{Classical definition of the signal-to-noise ratio}
\label{sec:classical_snr}
The point of this section is to show that the definition of an anomalous signal-to-noise ratio in Section~\ref{sec:scoring_rules} is the natural generalisation of the classical definition as given by various authors.
This classical definition refers to the {\em ratio of predictable components} or $\RPC$, and the signal-to-noise ratio is defined as anomalous if $\RPC > 1$.
We will show that the generalised and the classical definitions agree under the following circumstances (to be made precise below): 
\begin{enumerate}
    \item For the scoring rule in Definition~\ref{def:snp} we use the mean-squared error of the forecast mean.
    \item The forecast means of $f$ and of $\pi$ are in a linear relationship, or in other words, the mean-square optimal regression of the forecast mean onto the verification is linear.
\end{enumerate}
To properly state our results, we define for any probability distribution $\mu$ on $E$ the 
\beq{eq:1.50}\begin{split}
        \text{Mean} \qquad m(\mu) & := \int x \mu(x) \idd x \\
        \text{Variance} \qquad v(\mu) & := \int x^2 \mu(x) \idd x - m(\mu)^2.
\end{split}\eeq
In ``ensemble forecast'' terms, $m(f)$ is the ensemble mean, $v(f)$ the ensemble spread, and $\E(m(f)^2)$ the variability of the ensemble mean, sometimes referred to as the ``signal''.
We stress that if $\mu$ is a probability forecast (of whatever nature), then $m(\mu)$ and $v(\mu)$ will be random variables. 
\begin{defnt}\label{def:classic_snp}
Let $f$ be a probability forecast and $Y$ a verification.
The classical ratio of predictable components~\citep[see, e.g.,][]{weisheimer_confident_winter_NAO_2019} is defined as
\beq{eq:1.60}
\RPC = \frac{r(m(f),Y)}{r(m(f),Z)},
\eeq
where $r(m(f),Y)$ is the Pearson correlation coefficient between $m(f)$ and the observation $Y$, while $r(m(f), Z)$ is the Pearson correlation coefficient between $m(f)$ and a random variable $Z$ drawn from the forecast $f$.
The forecast $f$ exhibits an {\em anomalous signal-to-noise ratio} if $\RPC > 1$.
\end{defnt}
Broadly speaking, the definition states that the forecast $f$ exhibits an anomalous signal-to-noise ratio if $m(f)$ correlates better with the verification than with draws from $f$.
We also note that the denominator in Equation~\eqref{eq:1.60} can be written as $\sqrt{\E(m(f)^2) / v(\meanf)}$.
As announced, we will now clarify the connection between two notions of an anomalous signal-to-noise ratio as in Definitions~\ref{def:snp} and~\ref{def:classic_snp}.
\begin{thm}\label{thm:SNP_linear}
Let $f$ be a probability forecast and $Y$ a verification, and let $\pi$ be the recalibration of $f$ with respect to $Y$.
Suppose that 
\begin{enumerate}
\item $\E Y = m(\meanf) = 0$,
\item\label{itm:linear} $m(\pi) = a + b \, m(f)$, with $a, b$ determined so as to minimise $\E((Y - m(\pi))^2)$.
\end{enumerate}
Then $\RPC \geq 1$ if and only if $\RSS_{\sigma} \geq 1$, where $\RSS_{\sigma}$ is understood with respect to the specific scoring rule $\sigma(f, y) := (y - m(f))^2$.
In particular the two notions of an anomalous signal-to-noise ratio (from Defs.~\ref{def:snp} and~\ref{def:classic_snp}) agree in this case.
\end{thm}
A few remarks are in order:
\begin{enumerate}
\item Due to the definition of $\pi$ we have $m(\pi) = \E(Y|f)$, so Assumption~2 in Theorem~\ref{thm:SNP_linear} amounts to saying that $\E(Y|f)$, which will generally be a nonlinear function of $f$, is actually a linear function of $m(f)$.
\item The scoring rule $\sigma(f, y)$ is indeed proper but not strictly proper.
\item Assumption~\ref{itm:linear} in the Theorem formalises the informal Assumption~2 stated at the beginning of the present Section.
\item Although both $\RPC = 1$ and $\RSS_s = 1$ have the interpretation of a normal SNR, their magnitudes cannot be directly compared for an anomalous SNR, not even for the specific score $\sigma$; in general $\RSS_{\sigma}$ will not be numerically equal to the classical $ \RPC$.
\end{enumerate}
We will now sketch the proof of Theorem~\ref{thm:SNP_linear}.
Some details will be given in Appendix~\ref{sec:appendix}.
First we note that 
\beq{eq:1.67}
r(m(f), Y) = r(m(\pi), Y),
\eeq
simply because $m(\pi)$ is a linear function of $m(f)$ due to Assumptions~1, 2 in the theorem.
Next it follows from the definition of the correlation coefficient $r$ and of $m(\pi)$ that 
\beq{eq:1.69}
r(m(\pi), Y) 
= \frac{\E(m(\pi)Y)}{\sqrt{\E(Y^2)\; \E(m(\pi)^2)}}
= \frac{\E(m(\pi)^2)}{\sqrt{v(\meanpi)\; \E(m(\pi)^2)}}
= \sqrt{\frac{\E(m(\pi)^2)}{v(\meanpi)}},
\eeq
and the same expression holds for $r(m(f), Z)$ when replacing $f$ for $\pi$ and $Z$ for $Y$ as was remarked already just after Definition~\ref{def:classic_snp}. 
Thus we have that
\beq{eq:1.71}
\RPC^2 = \frac{\E(m(\pi)^2) / v(\meanpi)}{\E(m(f)^2) / v(\meanf)}.
\eeq
Turning to the explicit form of $\SSC$ in case of the scoring rule $\sigma$, one finds (see Appendix~\ref{sec:appendix}) that
\beq{eq:1.73}
\SSC(f) = \frac{v(\meanf) - \E(m(f)^2)}{v(\meanf)}
= 1 - \frac{\E(m(f)^2)}{v(\meanf)},
\eeq
and the same for $\pi$. 
Comparing Equations~\eqref{eq:1.71} and~\eqref{eq:1.73}, we find that $\RPC^2 > 1 $ if and only if $\RSS_{\sigma} > 1$, as claimed.
This finishes the proof of Theorem~\ref{thm:SNP_linear}.%

We conclude from this theorem that by replacing $\RSS_{\sigma}$ for $\RPC$ we obtain an entirely equivalent definition of an anomalous signal-to-noise ratio.
Yet our interpretation of $\RSS_{\sigma}$ applies just as well if we replace $\sigma$ with any other proper scoring rule $s$. 
We can therefore interpret the corresponding definitions of an anomalous signal-to-noise ratio as generalisations of the classical definitions.
\section{An anomalous signal-to-noise ratio---so what?} \label{sec:RPC_interpretation}
It is often stated, somewhat imprecisely, that an anomalous signal-to-noise ratio implies that the ensemble mean is better at forecasting the actual observation than it is at forecasting the individual ensemble members. 
It must be kept in mind though that the performance is ``better'' only in a relative sense, since our generalised formulation of an anomalous signal-to-noise ratio (which includes the classical definition) is in terms of {\em skill~scores}, that is of performance {\em relative} to the predictive performance of climatology.
It would therefore be more accurate to describe the classical anomalous signal-to-noise ratio as a situation in which the error between the recalibrated ensemble mean and the observation, relative to the variance of the observation, is lower than the error between the ensemble mean and the ensemble members, relative to the variance of the ensemble members. 
This is also evident from the characterisation~\eqref{eq:1.71} of the classical~$\RPC$ in the proof of Theorem~\ref{thm:SNP_linear}.
We can conclude that since the definition of an anomalous signal-to-noise ratio refers to the skill score rather than the score itself, anomalous signal-to-noise ratios might occur because either the score of the recalibrated forecast against actual observations is better than expected, or because the climatology of the actual observations has a larger entropy than expected, or a combination of both; see Equation~\eqref{eq:1.41}.
In general, these two issues cannot be disentangled without further information or assumptions.
For instance we may assume (or better still, confirm statistically) that the overall score of $f$ against actual observations is equal to what we called the self-score of $f$, that is the score of $f$ against hypothetical observations drawn from $f$ itself~\citep[see, e.g.,][where this assumption is made]{brocker_statistical_2023,weisheimer_confident_winter_NAO_2019}. 
For the quadratic scoring rule $\sigma$ this would mean
\beq{eq:1.800}
\E(v(f)) = \E((Y - m(f))^2).
\eeq
For ensembles, this relation states that the ensemble spread is equal to the ensemble mean error. 
The relation between these two quantities is well known as the {\em spread--skill relationship}.
For the right-hand side however we have
\beq{eq:1.80a}
\E((Y - m(f))^2) = \E(v(\pi)) +  \E((m(f) - m(\pi))^2),
\eeq
which follows from a direct calculation.
Since the second term on the right-hand side of Equation~\eqref{eq:1.80a} is nonnegative, we find by combining Equations~\eqref{eq:1.80a} and~\eqref{eq:1.800} that
\beq{eq:1.85}
\E(v(f)) \geq  
\E(v(\pi)).
\eeq
In ensemble terms, the conclusion is that under a perfect spread--skill relationship, the spread of the original ensemble is at least as large as the spread of the recalibrated ensemble.
The relation~\eqref{eq:1.85} combined with a {\em normal} signal-to-noise ratio allows one to conclude that 
\beq{eq:1.90}
\E(m(f)^2) \geq 
\E(m(\pi)^2), 
\eeq 
as is easily seen, meaning that the actually achievable signal of $f$ (through proper recalibration) is smaller than the anticipated signal of $f$ (i.e.,\ the signal we would see if verifications were drawn from $f$).
However, if the signal-to-noise ratio is {\em anomalous}, in contrast, the converse relation to~\eqref{eq:1.90} does {\em not} necessarily hold!
It follows directly from the definition of $\SSC$ that there are two ways in which the signal-to-noise ratio can be anomalous: either because the climatology of the actual observations has a larger variance than the averaged forecast, that is $v(\meanf) \leq v(\meanpi)$, or because the score of the recalibrated forecast against actual observations is better than expected, that is $\E(v(f)) \geq \E(v(\pi))$, or both.
The latter case however is already guaranteed by relation~\eqref{eq:1.85}, so it is impossible to decide whether $v(\meanf)$ is larger or smaller than $v(\meanpi)$ in this situation, at least without further data analysis.
We conclude that since the $\RPC$ is a ratio of skill scores and thus a measure of relative performance, it needs to be complemented with other information in order to draw conclusions about absolute performance.
One possibility is to assume (or statistically confirm) a perfect spread--skill relationship.
%
Whether and how the spread--skill relationship can be generalised so as to complement the $\RSS$ will be subject to future research.
\section{Numerical experiments}\label{sec:numerical_experiments}
In this section, we discuss several numerical experiments to compare estimators of the new $\RSS$ concept for different scoring rules and forecast types, including the classical $\RPC$, for ensemble forecasts.
In Section~\ref{ssec:synthetic} we present experiments using synthetic data with known (classical) $\RPC$, that is, the $\RPC$ can be controlled in these data.
Subsequently in Section~\ref{ssec:realdata}, we will consider seasonal hindcasts and reanalysis of the North Atlantic Oscillation (NAO) index.
Both data sets comprise ensemble forecasts and corresponding real-valued verifications.
This allows the application of the classical $\RPC$.
Furthermore, we will consider the $\RSS$ for the continuous ranked probability score (CRPS), a probabilistic scoring rule that can be applied directly to real-valued ensembles and the corresponding verifications~(see Section~\ref{sssec:crps}). 
We also consider binary verifications and corresponding probability forecasts; these will be generated from the real-valued forecasts discussed earlier by thresholding. 
The logarithmic score will be used for those experiments~(see Section~\ref{sssec:logarithmic_sr}).
Note that we will obtain different versions of the $\RSS$, depending on which score is used, yet for the same forecasting system. 
We expect that an anomalous signal-to-noise ratio in a forecasting system is a fundamental statistical property of that system which, if present, should be picked up by different scoring rules, amount of data permitting.
When analysing operational forecast data, one does not have access to the optimally recalibrated forecasts $\pi$, which is a theoretical construction based on certain mathematical assumptions.
Recall that $\pi$ is a (potentially highly complicated and nonlinear) function which maps instances of the forecast $f$ onto the actual probability distribution of the verification $Y$, conditional on the forecast $f$; it might be thought of as the optimal postprocessing of $f$.
Evidently, $\pi$ can only be approximated, based on the available data.
There is a plethora of probabilistic postprocessing methods available; see, for instance, \citet{vannitsem2018statistical} for an overview and further references regarding the postprocessing of ensembles.
The specific method selected for approximating $\pi$ will of course influence the statistical properties of the resulting estimator of $\RSS$, and may in fact affect whether or not one finds there to be an anomalous signal-to-noise ratio.
A detailed analysis of these properties, and in particular the influence of estimating $\pi$, will be the subject of future research.
In the following experiments, we will employ rather simple, and most likely suboptimal, approaches to estimating $\pi$.
The details will be discussed below.

Moreover, when $\pi$ is not the optimal recalibration, the values of the expressions for the RSS in Equations \eqref{eq:rss_s} and \eqref{eq:1.41} will differ. Here, we use Equation~\eqref{eq:rss_s}.
\subsection{Experimental setup}\label{ssec:experimental_setup}
In all the following experiments, we use bootstrap resampling~\cite{efron_computer_2016} across the time indices (that is, the forecast--verification pairs for each time are sampled with replacement) to approximate the sampling distribution of each statistic, in particular our estimates of the $\RPC$ and $\RSS$. This was also done in \citet{strommen_relationship_2023}.

We use the \verb|scoringrules| Python library \citep{zanetta_scoringrules_2024} for computing scoring rules and the SciPy library \citep{virtanen_scipy_2020} for optimisation (see below for more details).
The open-source Python code for the experiments performed here is available at \url{https://github.com/eviatarbach/snp}.
In the following two subsections we discuss the scoring rules and corresponding $\RSS$ to be used for the ensemble forecasts with real-valued verifications, and for the probability forecasts with binary verifications. 
Furthermore, we provide details as to how we approximate the optimally calibrated forecast $\pi$.
\subsubsection{Ensemble forecasts and the CRPS}\label{sssec:crps}
We consider verifications $\{y_n; n = 1, \ldots, N\}$ and corresponding ensemble forecasts $\{x^{(k)}_n; n = 1, \ldots, N, \; k = 1, \ldots, K\}$, both consisting of real numbers. 
We will write the time index as a subscript, while superscripts enumerate the ensemble members. 
Consequently, $N$ is the number of forecast--verification pairs in our data set, while $K$ is the number of ensemble members. 
To score ensemble forecasts we use the {\em continuous ranked probability score}~\cite[CRPS; see][]{brown_admissible_1974,wilks_statistical_2019}, a strictly proper scoring rule.
In its original form, the CRPS is defined for a real-valued verification $y$ and a forecast $f$ given as a cumulative distribution function (CDF), and can be written as 
\begin{equation}
    \operatorname{CRPS}(f, y) = \int (f(x) - H(x - y))^2 \idd x,
\end{equation}
where $H$ is the Heaviside step function.
This scoring rule has become popular in meteorological forecast verification, since it (and its multidimensional extensions) are fairly easy to evaluate, in particular for ensemble forecasts, as we will now see.
An ensemble forecast $\{x^{(k)}\}_{1 \leq k \leq K}$ can be transformed into a piecewise constant CDF $f$ through
\beq{eq:3.10}
    f(x) = \frac{1}{K} \sum_{k=1}^{K} \mathbb{1}\{x^{(k)} \leq x\},
\eeq
where $\mathbb{1}$ is the indicator function.
Subsequently, we take the CRPS of an ensemble $\{x^{(k)}\}_{k \leq K}$ to be the~CRPS of the corresponding~CDF defined through Equation~\eqref{eq:3.10}.
(We will also write $f$ to refer to the ensemble since the CDF and the ensemble convey exactly the same information.)
In order to compute the $\RSS$ we need to find $\pi$ defined in Equation~\eqref{eq:1.20}.
We recall that mathematically speaking, $\pi$ is a function which maps the original forecast $f$ onto another forecast which is perfectly calibrated.
In the present case, this means that $\pi$ transforms the original ensemble onto a perfectly calibrated ensemble.
As discussed, we do not have direct access to $\pi$, so we need to approximate it by fitting a parametric model to the data.
We will restrict ourselves here to a very simple linear approach. 
More specifically, we shift the mean of $f$ according to $m(\pi) = a + b \, m(f)$ while leaving the ensemble spread around the mean the same.
We determine $a$ and $b$ so as to minimise the averaged CRPS (over all forecast--verification pairs) of the transformed ensemble, namely~$\sum_{n=1}^N \operatorname{CRPS}(\pi_n, y_n)$.
The minimisation is carried out using the Broyden--Fletcher--Goldfarb--Shanno~(BFGS) optimisation method.
Note that this approach is ``ensemble--aware'' in that the resulting ensemble is still exchangeable if the original ensemble was exchangeable~\citep{broecker_exchangeability_2011}. See also \citet{brener_hypothesis_2024} for a discussion of the importance of exchangeability in the context of the signal-to-noise paradox.
\subsubsection{Binary probability forecasts and the logarithmic scoring rule}\label{sssec:logarithmic_sr}
In binary forecasting problems, the verification $Y$ can only take the values $0$ or $1$ (or more generally two arbitrarily labelled categories).
Probability forecasts for such problems provide a number $f \in [0, 1]$ representing the forecast probability of the outcome $Y = 1$, while $1-f$ is the probability of outcome $Y = 0$.
Such forecasts can be scored using the {\em logarithmic scoring rule}~\cite{good_rational_1952,wilks_statistical_2019}, a proper scoring rule given by the formula
\beq{eq:3.20}
    \operatorname{LS}(f, Y) = -\log|p + Y - 1|.
\eeq
Again we need to find $\pi$, which in the present case is a function from the unit interval to the unit interval.
(As mentioned earlier, readers familiar with the reliability diagram may visualise $\pi$ as the perfect recalibration curve.)
We approximate $\pi$ through a logit model
\beq{eq:3.30}
    \operatorname{logit}(\pi) = a + b \operatorname{logit}(f),
\eeq
with parameters $a, b$, where $\operatorname{logit}$ is the logit transformation
\begin{equation}
    \operatorname{logit}(p) = \log\left(\frac{p}{1 - p}\right).
\end{equation}
We choose $a$ and $b$ to minimise the total logarithmic score $\sum_{n=1}^N\operatorname{LS}(\pi(f_n), y_n)$ over the forecast--verification archive.
The logarithmic score is infinite in case that $Y = 1$ yet $\pi = 0$, or if $Y=0$ yet $\pi = 1$.
Now $\pi = 0$ or $1$ happens if and only if $f = 0$ or $1$, respectively, irrespective of the values of $a, b$.
In order for this problem not to spoil the optimisation, we set $f_n = \epsilon$ in case $f_n = 0$, and $f_n = 1 - \epsilon$ in case $f_n = 1$, and we set $\epsilon = 10^{-2}$.
We note that in our experiments, the possible values of the forecast $f_n$ are discrete, and in particular values between $0$ and $\epsilon$ or between $1 - \epsilon$ and $1$ cannot occur (if we set $\epsilon$ sufficiently small).
\subsection{Simulations for synthetic data}\label{ssec:synthetic}
In this section, we discuss numerical experiments using synthetic data with known characteristics.
In particular, we are able to specify the value of the true $\RPC$, and thus whether the data exhibits an anomalous signal-to-noise ratio or not. 
These data comprise ensemble forecasts and corresponding verifications; we will also consider binary verifications with corresponding probability forecasts derived from these data.
To generate the data, we first draw a verification $Y$ from $\mathcal{N}(m_Y, \sin(\phi)^2)$, where $\phi$ is prescribed.
The mean $m_Y$ is itself drawn from $\mathcal{N}(0, \cos(\phi)^2)$. As a result, $Y$ has overall variance equal to one.
The forecast $f$ is defined by an ensemble drawn from $\mathcal{N}(m_f, \sigma_f^2)$. The mean $m_f$ is set to be $m_f = c m_Y$, where $c$ is a prescribed parameter.
The variance (or ensemble spread) $\sigma_f^2$ is set so as to match the expected error in the ensemble mean, that is we define $\sigma_f^2 := \E((m_f - Y)^2) = \sin(\phi)^2 + (1-c)^2\cos(\phi)^2$.
It is easy to see that $\cos(\phi) = r(m_Y, Y)$ is the correlation between $m_Y$ and the verification $Y$ (and is furthermore equal to $r(m_f, Y)$).
We define $\psi$ so that $\cos(\psi) = r(m_f, Z)$ is the correlation between the mean $m_f$ of $f$ and a draw $Z$ from $f$. 
Consequently, we have that the $\RPC = \cos(\phi)/\cos(\psi)$. Note, however, that any specific generated instance of the synthetic verification and ensembles will generally have an empirical $\RPC$ that differs from this analytical value.
From the ensemble, we also form binary forecasts and verifications.
As verification we use the label ``1'' if $Y > 0$ and ``0'' else. 
To form the forecast $f$, we transform the ensemble members into binary labels in the same fashion and then use as the probability forecast the empirical frequency of the label ``1'' among the ensemble members.
\subsubsection{Normal signal-to-noise ratio}
We first experiment with the synthetic data in a situation where the true $\RPC = 1$. 
As further parameters we set $\phi = 0.3\pi$, and use $K = 25$~ensemble members for $f$. 
The length of the forecast--verification archive is $N = 100$ (larger than typical archives of seasonal forecasts, but comparable to subseasonal forecast archives).
From these data, we compute the empirical classical $\RPC$ as well as $\RSS_{\text{CRPS}}$ and $\RSS_{\text{LS}}$.
To obtain estimates of the variability of these statistics, we generate bootstrap resamples of the forecast--verification archive by drawing 100~samples with replacement from the original archive.
The results are summarised in Table~\ref{tab:normal}. 
\begin{table}
    \centering
    \begin{tabular}{lllll}
        & Non-resampled data & Bootstrap 2.5\%  & Bootstrap 50\% & Bootstrap 97.5\% \\
        Classical RPC & 1.01 & 0.84 & 1.00 & 1.11 \\
        $\RSS_\text{CRPS}$ & 0.99 & 0.90 & 1.00 & 1.10 \\
        $\RSS_\text{LS}$ & 0.89 & 0.83 & 0.90 & 1.04
    \end{tabular}
    \caption{The empirical values of the classical $\RPC$, $\RSS_{\text{CRPS}}$, and $\RSS_{\text{LS}}$ for the synthetic data with normal signal-to-noise ratio ($\RPC = 1$), along with quantiles of the bootstrap distribution.}
    \label{tab:normal}
\end{table}
The ratios of skill scores for both the CRPS and the binary logarithmic score fail to give a strong indication of an anomalous $\SNR$: a value of $\RSS = 1$ is within the 95\% bootstrap confidence intervals for both of these scoring rules.
This was to be expected as the forecast is, by design, reliable, and should not exhibit any $\SNR$~problem.
The experiment thus confirms that the methodology is working as intended and the $\RSS$s defined through scoring rules provide values in the expected range.
An interesting aspect of the results is the different levels of variation exhibited by the different $\RSS$s as shown by the intervals between the 5\% and 95\% bootstrap quantiles.
In particular, we see that $\RSS_{\text{CRPS}}$ has the smallest variability, followed by $\RSS_{\text{LS}}$, and finally the classical $\RPC$, with the largest variance. 
This indicates that $\RSS$s corresponding to different scoring rules exhibit different levels of variability and potentially statistical robustness. 
Therefore, depending on the problem at hand, $\RSS$s based on particular scoring rules might exhibit better statistical properties than the classical $\RPC$ and hence lead to superior methods of analysing forecasting systems.
Clearly, more research is needed in this direction, in particular a closer analysis of the statistical properties of the $\RSS_s$ for particular scoring rules $s$.
It was already shown in~\citet{brocker_statistical_2023}, for instance, that the variability of the classical $\RPC$ increases with decreasing correlation between ensemble mean and verification, which in the case of the synthetic data corresponds to a larger angle $\phi$. 
Presumably more relevant, however, is the ability of the diagnostic statistics to distinguish between cases of normal and anomalous signal-to-noise ratios, and we will come back to this in the next subsection where we analyse data with a true $\RPC$ different from unity. 
\subsubsection{Anomalous signal-to-noise ratio}\label{ssec:anomalous_synthetic}
Next we analyse synthetic data with an anomalous signal-to-noise ratio of $\RPC = 1.52$.
This corresponds to $c = 0.6$, and we let $\phi = 0.3\pi$ (as before).
We still consider 25~ensemble members and 100~samples.
The results are summarised in Table~\ref{tab:anomalous}. 
\begin{table}
    \centering
    \begin{tabular}{lllll}
        & Non-resampled data & Bootstrap 2.5\%  & Bootstrap 50\% & Bootstrap 97.5\% \\
        Classical RPC & 1.54 & 1.20 & 1.55 & 1.82 \\
        $\RSS_\text{CRPS}$ & 1.17 & 1.08 & 1.17 & 1.29 \\
        $\RSS_\text{LS}$ & 1.09 & 1.05 & 1.18 & 1.36
    \end{tabular}
    \caption{The empirical values of the classical $\RPC$, $\RSS_{\text{CRPS}}$, and $\RSS_{\text{LS}}$ for the synthetic data with anomalous signal-to-noise ratio ($\RPC = 1.52$), along with quantiles of the bootstrap distribution.}
    \label{tab:anomalous}
\end{table}
For both ratios of skill scores (corresponding to the logarithmic and CRPS~score), a value of~1 is outside the 95\% confidence interval.
Specifically, the ratios of skill scores are typically larger than~1 for both the~CRPS and the logarithmic score. 
Hence, the $\RSS$ for both scores provides strong indication of an anomalous signal-to-noise ratio.
\begin{figure}
    \centering
    \begin{subfigure}{0.45\textwidth}
        \includegraphics[width=\linewidth]{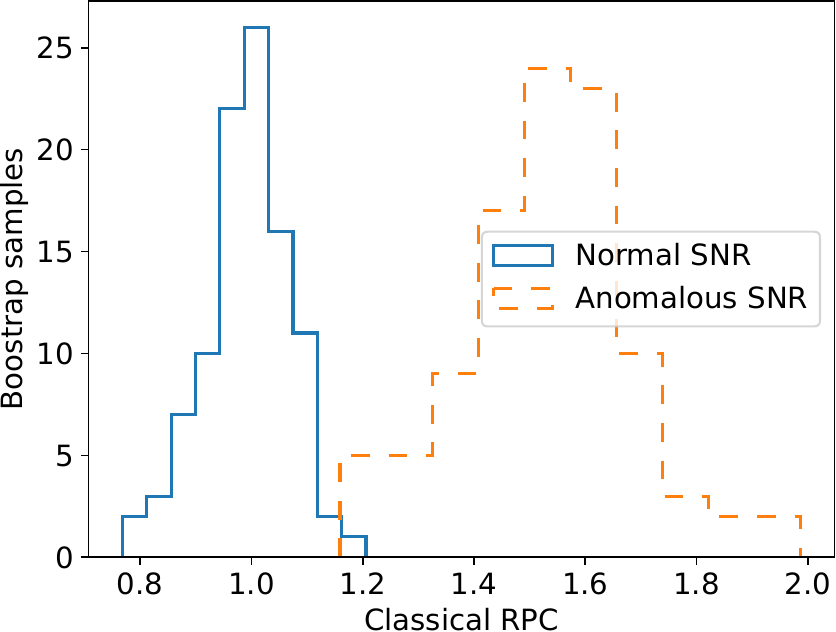}
        \caption{}
    \end{subfigure}\hfill
    \begin{subfigure}{0.45\textwidth}
        \includegraphics[width=\linewidth]{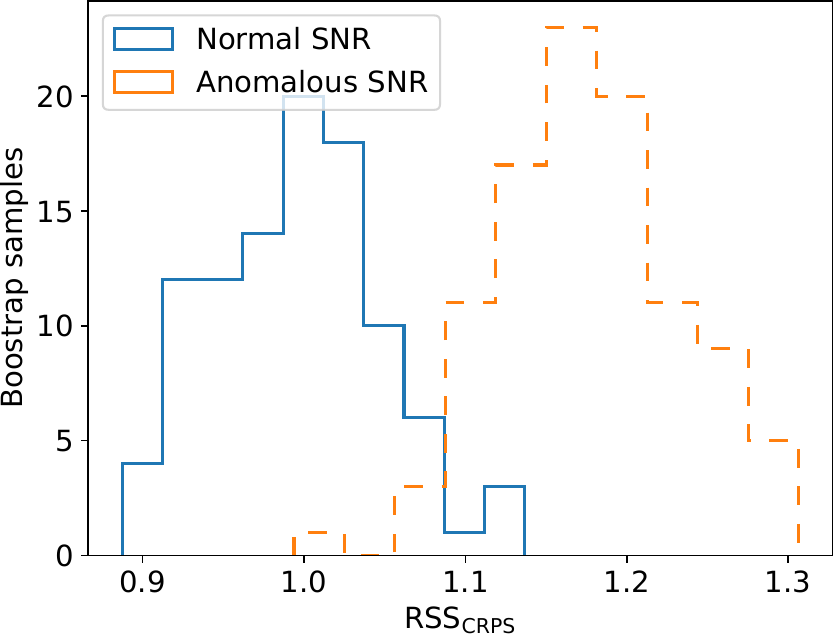}
        \caption{}
    \end{subfigure}
    \begin{subfigure}{0.45\textwidth}
        \includegraphics[width=\linewidth]{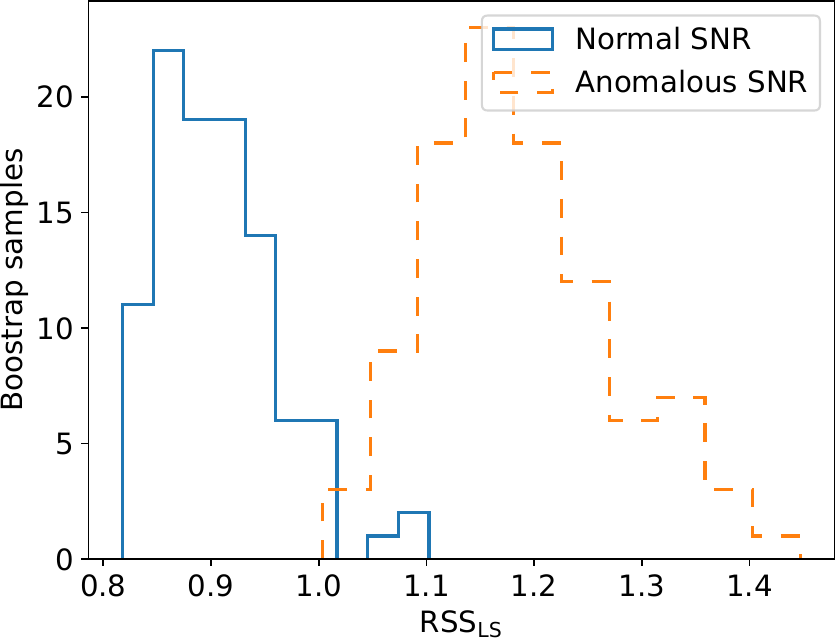}
        \caption{}
    \end{subfigure}
    \caption{Histograms of bootstrap samples of (a) classical RPC (b) $\RSS_\text{CRPS}$ and (c) $\RSS_\text{LS}$ for a normal SNR and an anomalous SNR.}
    \label{fig:bootstrap}
\end{figure}
Figure~\ref{fig:bootstrap} contains another way of visualising these results, showing the bootstrap distributions for both the normal and anomalous $\SNR$ cases, and for both $\RSS_\text{CRPS}$ and $\RSS_\text{LS}$.
A marked rightward shift of the bootstrap distributions can be seen between the normal and anomalous cases, indicating that both scores provide strong evidence of an anomalous $\SNR$.
It needs to be stressed, however, that the ability of the $\RSS$ or $\RPC$ statistics to distinguish between cases of normal and anomalous signal-to-noise ratios depends on their statistical properties under {\em both} scenarios.
This is illustrated by comparing the results in Table~\ref{tab:normal} with those in Table~\ref{tab:anomalous}, or alternatively comparing the blue and yellow graphs in Figure~\ref{fig:bootstrap}.
A test based on the classical $\RPC$, for instance, could declare the signal-to-noise ratio to be anomalous if $\RPC > 1.11$.
From Table~\ref{tab:normal}, we find that the probability of false positives would be less than 2.5\%, while the probability of false negatives would be about the same by Table~\ref{tab:anomalous}.
For $\RSS_\text{CRPS}$ and $\RSS_\text{LS}$ we get about the same numbers except that we have to take different thresholds, namely about~1.08 for $\RSS_\text{CRPS}$ and~1.05 for $\RSS_\text{LS}$.
Figure~\ref{fig:bootstrap} gives the same conclusion, except that probably lower false positive and false negative probabilities could be achieved by using $\RPC$ with a threshold of about $1.2$, although these numbers are based on too few samples for definite conclusions. 
We {\em can} conclude, however, that a decision threshold equal to~$1$ would not be a good idea for $\RPC$ or $\RSS_{\text{CRPS}}$ as it would give false positive rates of about 50\%. 
%
%
We also note that although none of the three statistics emerges as the ``winner'', all of them appear to contain similar information, and they use different forecast information between them---especially $\RSS_{\text{LS}}$, which applies to forecasts that are not accessible to the classical $\RPC$.
\subsection{Experiments for the North Atlantic Oscillation index}\label{ssec:realdata}
\begin{figure}
    \centering
    \includegraphics[width=1\linewidth]{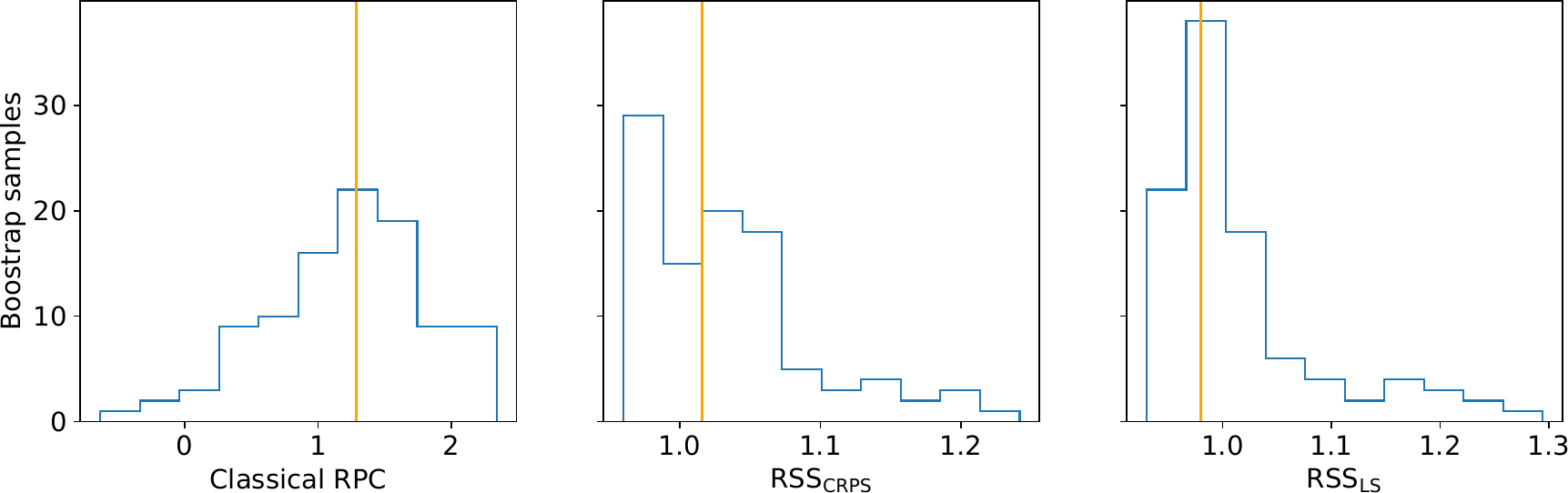}
    \caption{Histograms of bootstrap samples of classical RPC, $\RSS_\text{CRPS}$, and $\RSS_\text{LS}$ for the NAO forecasts. The orange vertical lines show the estimate obtained from the non-resampled data.}
    \label{fig:bootstrap_nao}
\end{figure}

We now apply the ratios of skill scores to ensemble hindcasts and reanalyses (as verifications) of the North Atlantic Oscillation~(NAO) index. More specifically, we use the first empirical orthogonal function of the 500 hPa geopotential height field, averaged over the months December, January, and February, as is used routinely at the European Centre for Medium-Range Weather Forecasts (ECMWF).
The forecasts are taken from the~SEAS5 seasonal hindcast data set \cite{johnson_seas5_2019}, using a 25-member ensemble initialised on November~1st each year during the hindcast period of 1981--2009.
We take the verification data to be the ERA5~reanalysis, ECMWF’s latest atmospheric reanalysis \cite{hersbach_era5_2020} product.\footnote{Reanalyses are obtained by assimilation of historical observations into a weather forecasting model, and thus strictly speaking do not consist of direct observations of atmospheric variables.}
Table~\ref{tab:nao} and Figure~\ref{fig:bootstrap_nao} summarise, respectively, the bootstrap distributions for the classical $\RPC$, as well as $\RSS_\text{CRPS}$ and  $\RSS_\text{LS}$, calculated for this data using the methodology described above.
\begin{table}
    \centering
    \begin{tabular}{lllll}
        & Non-resampled data & Bootstrap 2.5\%  & Bootstrap 50\% & Bootstrap 97.5\% \\
        Classical $\RPC$ & 1.29 & $-0.07$ & 1.32 & 2.20 \\
        $\RSS_\text{CRPS}$ & 1.02 & 0.97 & 1.03 & 1.20\\
        $\RSS_\text{LS}$ & 0.98 & 0.95 & 0.99 & 1.23
    \end{tabular}
    \caption{The values of the classical RPC and $\RSS_\text{CRPS}$ for the NAO index forecasts, along with quantiles of the bootstrap distribution.}
    \label{tab:nao}
\end{table}
For the classical RPC we obtain a bootstrap median larger than~1. However, the bootstrap median of $\RSS_\text{CRPS}$ and $\RSS_\text{LS}$ is close to 1.
An $\RPC$ of~1 lies within the 95\% confidence intervals of all three bootstrap distributions.
Therefore, based on this limited amount of data we do not find evidence for an anomalous signal-to-noise ratio.
This conclusion is consistent across the $\RPC$ and the two $\RSS$s.
However, we observe that $\RSS_\text{CRPS}$ has a much lower variance than the classical $\RPC$, namely about~0.2 compared to about~1.9.
This means that the classical $\RPC$ contains very little information, while $\RSS_\text{CRPS}$ provides much clearer evidence that the verification is approximately as predictable as the ensemble members.
\section{Conclusions}\label{sec:conclusions}
A generalised concept of the signal-to-noise ratio is presented, based on proper scoring rules rather than merely linear correlations.
The guiding idea is to define the signal-to-noise ratio as the ratio of two {\em skill scores}, namely the skill score of the forecast against a verification drawn from the forecast, versus the skill score of the perfectly recalibrated forecast against the actual verifications.
It is shown that this definition is the natural generalisation of the classical ratio of predictable components ($\RPC$), but it can be applied to any type of forecast that is amenable to scoring, not only real-valued ensemble forecasts and verifications.
This drastically widens the applicability of these concepts.
As examples, we consider ensemble forecasts, scored using the continuous ranked probability score~(CRPS), and probability forecasts of a binary event, scored using the logarithmic score. 
The numerical examples show that the new ratio of skill scores ($\RSS$) basically delivers the same message as the classical $\RPC$, thus demonstrating that it can indeed be considered a suitable generalisation of the classical $\RPC$.
At the same time, however, the experiments already indicate that the variance of the different statistics can be very different. 
For instance, in the context of an actual seasonal hindcast data set (of the North Atlantic Oscillation index), the $\RSS$ based on the CRPS exhibits a variance which is a mere 10\% of the classical $\RPC$'s variance.
This suggests that depending on the problem at hand, a ratio of skill scores potentially exhibits much better statistical properties than the classical $\RPC$, and thus permits more confident statements as to whether a given forecasting system exhibits an anomalous signal-to-noise ratio or not, despite using the same data.
However, as was stressed in Section~\ref{ssec:anomalous_synthetic}, the ability of the $\RSS$ or $\RPC$ statistics to distinguish between cases of normal and anomalous signal-to-noise ratios is not just a matter of the variance of those statistics (under either one or both scenarios).
To be suitable as a test statistic, the value of the $\RPC$ (or $\RSS$) statistic has to be significantly {\em different} between normal versus anomalous signal-to-noise~ratios.
Thus, the statistical properties of these quantities need to be investigated under {\em both} scenarios.
A future area of work will therefore be the analysis of the sampling properties of scoring rule--based $\RSS$s.
(In fact, this analysis is arguably incomplete even in the case of the classical $\RPC$, although see~\citet{brocker_statistical_2023} for an analysis linking the variance of the classical $\RPC$ to the correlation between ensemble mean and verification, which is comparatively weaker in seasonal forecasts as opposed to medium-range forecasts.)

A second area of future research is to consider more powerful methods for estimating $\pi$, the optimally recalibrated forecast.
In the present work we restricted ourselves to very simple linear recalibration (with a logit link in the case of forecast probabilities), which is probably warranted given the limited amounts of data.
It seems clear, however, that there is large scope here for methods of machine learning to obtain better estimates of $\pi$.

A third area of future research will be the investigation of possible dynamical mechanisms leading to anomalous signal-to-noise ratios, following previous work and open questions laid out in \citet{weisheimer_signal--noise_2024}, using the new statistical tools introduced herein.

\section*{Acknowledgments}

We would like to thank two anonymous reviewers, the Associate Editor, and Cadan Plasa for a number of helpful comments and suggestions.
These resulted in significant improvements of the paper. We thank Antje Weisheimer for providing the NAO hindcasts and verification data.
The work of both authors was partially funded by the AUSPICE project, UKRI2165.

\appendix

\section{Proof of Equation~(\ref{eq:1.25})}\label{sec:appendix1}
We would like to show that
$$    \E(s(f, Y)) = \E(\cS(f, \pi)).$$%
This follows from
\beq{eq:1.26}
        \E(s(f, Y)) = \E( \E(s(f, Y) | f) ) = \E\left(\int s(f, y) \pi(y) \idd y\right) = \E(\cS(f, \pi)).
\eeq

\section{Proof of Equation~(\ref{eq:1.73})}\label{sec:appendix}
We need to prove that for the scoring rule $\sigma(f, y) := (y - m(f))^2$ we have
\beq{eq:A.73}
\SSC(f) = \frac{\E(\cE(f))}{\cE(\meanf)}
= \frac{v(\meanf) - \E(m(f)^2)}{v(\meanf)},
\eeq
the first equality just being the definition of $\SSC$.
Now 
\beqn{eq:A.80}\begin{split}
\cE(f) 
& = \int \sigma(f, y) f(y) \idd y, \\
& = \int (y - m(f))^2 f(y) \idd y, \\
& = \int (y^2 - 2 y m(f)) f(y) \idd y + m(f)^2, \\
& = \int y^2 f(y) \idd y - m(f)^2. 
\end{split}\eeq
 Taking the expectation value $\E(\ldots)$, we find
$\E(\cE(f)) = \int y^2 \meanf(y) \idd y - \E(m(f)^2)$. 
But $\int y^2 \meanf(y) \idd y = v(\meanf)$, showing that $\E(\cE(f)) = v(\meanf) - \E(m(f)^2)$. 
Along similar lines it can be shown that 
$\cE(\meanf) = v(\meanf)$. 
Note that we assume here that $m(\meanf) = 0$ to simplify the calculations, but the statements remain correct if $m(\meanf) \neq 0$.
\bibliography{references}
\end{document}

%% file: macrofile.tex
\newcommand{\R}{\mathbb{R}}

\newcommand{\E}{\mathbb{E}}

\newcommand{\cD}{\mathcal{D}} 
\newcommand{\cE}{\mathcal{E}}

\newcommand{\cL}{\mathcal{L}}

\newcommand{\cS}{\mathcal{S}}

\newcommand{\meanpi}{\overline{\pi}}
\newcommand{\meanf}{\overline{f}}

\newcommand{\SSC}{\operatorname{SSS}}
\newcommand{\SNR}{\operatorname{SNR}}

\newcommand{\RPC}{\operatorname{RPC}}
\newcommand{\RSS}{\operatorname{RSS}}

\newcommand{\idd}{\; \mathrm{d}}
\newcommand{\beq}[1]{\begin{equation}\label{#1}}
\newcommand{\eeq}{\end{equation}}
\newcommand{\beqn}[1]{\begin{equation} \nonumber}
\newcommand{\labeln}[1]{ \nonumber}
\newtheorem{defnt}{Definition}
\newtheorem{thm}{Theorem}
\newcommand{\tinylbl}[1]{}
\makeatletter
\newcommand*{\add}{\@ifstar \@adds \@add}
\newcommand*{\addon}{\@ifstar \@addons \@addon}
\newcommand*{\del}{\@ifstar \@dels \@del}
\newcommand*{\delon}{\@ifstar \@delons \@delon}
\newcommand*{\edit}{\@ifstar \@edits \@edit}
\newcommand*{\mnote}{\@ifstar \@mnotes \@mnote}
\newcommand{\rmk}{\@rmk}
\ifthenelse{\boolean{@annotate}}{
\ifthenelse{\boolean{@draft}}{
\renewcommand{\tinylbl}[1]{\tiny{#1}}}{}
\RequirePackage{color}
\newcommand*{\@add}[2]{\linelabel{#1}\textcolor{blue}{\texttt{\tinylbl{#1}}#2}}
\newcommand*{\@addon}[1]{\linelabel{#1}\texttt{\tinylbl{#1}}\color{blue}}
\newcommand*{\@adds}[1]{\textcolor{blue}{#1}}
\newcommand*{\@addons}{\color{blue}}

\newcommand*{\@del}[2]{\linelabel{#1}\textcolor{red}{\texttt{\tinylbl{#1}}#2}}
\newcommand*{\@delon}[1]{\linelabel{#1}\texttt{\tinylbl{#1}}\color{red}}
\newcommand*{\@dels}[1]{\textcolor{red}{#1}}
\newcommand*{\@delons}{\color{red}}

\newcommand*{\@edit}[3]{\linelabel{#1}\textcolor{red}{\texttt{\tinylbl{#1}}#2} \textcolor{blue}{#3}}
\newcommand*{\@mnote}[2]{\linelabel{#1}\marginpar{\footnotesize\texttt{#2}}}
\newcommand*{\@edits}[2]{\textcolor{red}{#1} \textcolor{blue}{#2}}
\newcommand*{\@mnotes}[1]{\marginpar{\footnotesize\texttt{#1}}}
\newcommand{\@rmk}[1]{\par{\color{green} \texttt{#1}} \par}
}{%
\newcommand*{\@add}[2]{#2}
\newcommand*{\@addon}[1]{}
\newcommand*{\@adds}[1]{#1}
\newcommand*{\@addons}{}

\newcommand*{\@del}[2]{}
\newcommand*{\@delon}[1]{\color{red}}
\newcommand*{\@dels}[1]{}
\newcommand*{\@delons}{\color{red}}

\newcommand*{\@edit}[3]{#3}
\newcommand*{\@mnote}[2]{}
\newcommand*{\@edits}[2]{#2}
\newcommand*{\@mnotes}[1]{}
\newcommand{\@rmk}[1]{}
}
\makeatother
\ifthenelse{\boolean{@draft}}{%
\RequirePackage{showlabels}
\makeatletter
\def\SL@eqntext#1{\rlap{\quad\SL@margintext{#1}}}
\makeatother
\renewcommand{\beqn}[1]{\begin{equation}\label{#1}}%
}{}